\documentclass[english,aps,pra,reprint]{revtex4-1}   
\usepackage[T1]{fontenc}	
\usepackage[latin9]{inputenc}	
\usepackage{geometry}		
\usepackage[above,below]{placeins}	
\usepackage{times}
\usepackage{graphicx}
\graphicspath{ {images/} }
\usepackage[export]{adjustbox}
\usepackage{dcolumn}
\usepackage{bm}
\usepackage{enumitem}
\usepackage{color}
\usepackage[caption=false]{subfig}

\begin{document}


\title{Methods for Analyzing Pathways through a Physics Major}
\author{John M. Aiken}
\affiliation{Helmholtz Centre Potsdam GFZ German Research Centre for Geosciences, Potsdam, DE}

\author{Marcos D. Caballero}
\affiliation{Department of Physics and Astronomy \& CREATE for STEM Institute, Michigan State University, East Lansing, MI, USA}%

\begin{abstract}
Physics Education Research frequently investigates what students studying physics do on small time scales (e.g. single courses, observations within single courses), or post-education time scales (e.g., what jobs do physics majors get?) but there is little research into how students get from the beginning to the end of a physics degree. Our work attempts to visualize students paths through the physics major, and quantitatively describe the students who take physics courses, receive physics degrees, and change degree paths into and out of the physics program at Michigan State University.

\end{abstract}

\pacs{01.40.Fk, 01.40.G-}

\maketitle

\section{Introduction}\label{sec:intro}

Recruiting and retaining students in the physics major is an important challenge that departments across the country are facing \cite{chen2013,hodapbackpage}. Understanding the kinds of programs and practices that can support and sustain students intending to major or those currently majoring in physics is critical to grow a diverse population of physics graduates. The research that looks at specific student experiences to develop rich descriptions of how those experiences influence students' perceptions and choices provides some understanding \cite{irving2013,irving2015}. As does the work that uses prior student experiences to model eventual outcomes \cite{hazari2010}. Equally important is working to understand what might be learned using data from institutions themselves. For this project, we have collected student registration data at Michigan State University (MSU) in order to develop analytic methods that help unpack the pathways into and out of the major.

MSU has collected a wide body of data on students for the last 10+ years. This data set contains information on over 100,000 students who have taken math and physics courses at MSU. Two percent of these students have declared a physics major at some point in their academic career and 0.5\% of students have earned a bachelor's degree in physics. This data includes timestamped course and degree major choices, grades, and demographics such as gender, ethnicity, and family educational history. 

In this methods paper, we are interested in (1) understanding the means of analysis that provide information on students' paths into and out of the physics major, (2) developing visual representations of these analyses that communicate what paths students take through the major, and (3) describing a possible mechanism (inferred from the available data) that can explain what differentiates students who receive a degree in physics and those that do not. In doing this work, our aim is not to dismiss the rich work around retention and recruitment, e.g., Refs.~\citep{irving2013,irving2015,hazari2010}, but rather to provide additional context on that this (and other work) might draw. In this paper, we have not conducted an analysis using demographics.

\section{Michigan State Physics}\label{sec:msu}


%


MSU is a large, land grant university with approximately 39,000 undergraduate students currently enrolled. MSU has both a college of arts and sciences and a college of engineering and enrolls $>2000$ students in introductory physics courses annually. The student population is predominately white (65.7\%) with a sizable minority population (34.3\%). MSU has slightly more women enrolled than men (48\% men, 52\% women). The physics major enrolls a greater proportion white students (73.8\%) in comparison to the general MSU population and graduates a greater proportion as well (83.1\%). MSU physics graduate gender contrasts to the general population (83\% men, 17\% women) -- a proportion that is typical of physics departments across the country \cite{womeninphys}.

\section{Student Pathways}\label{sec:paths}

\begin{figure}[t]
\includegraphics[width=0.95\linewidth]{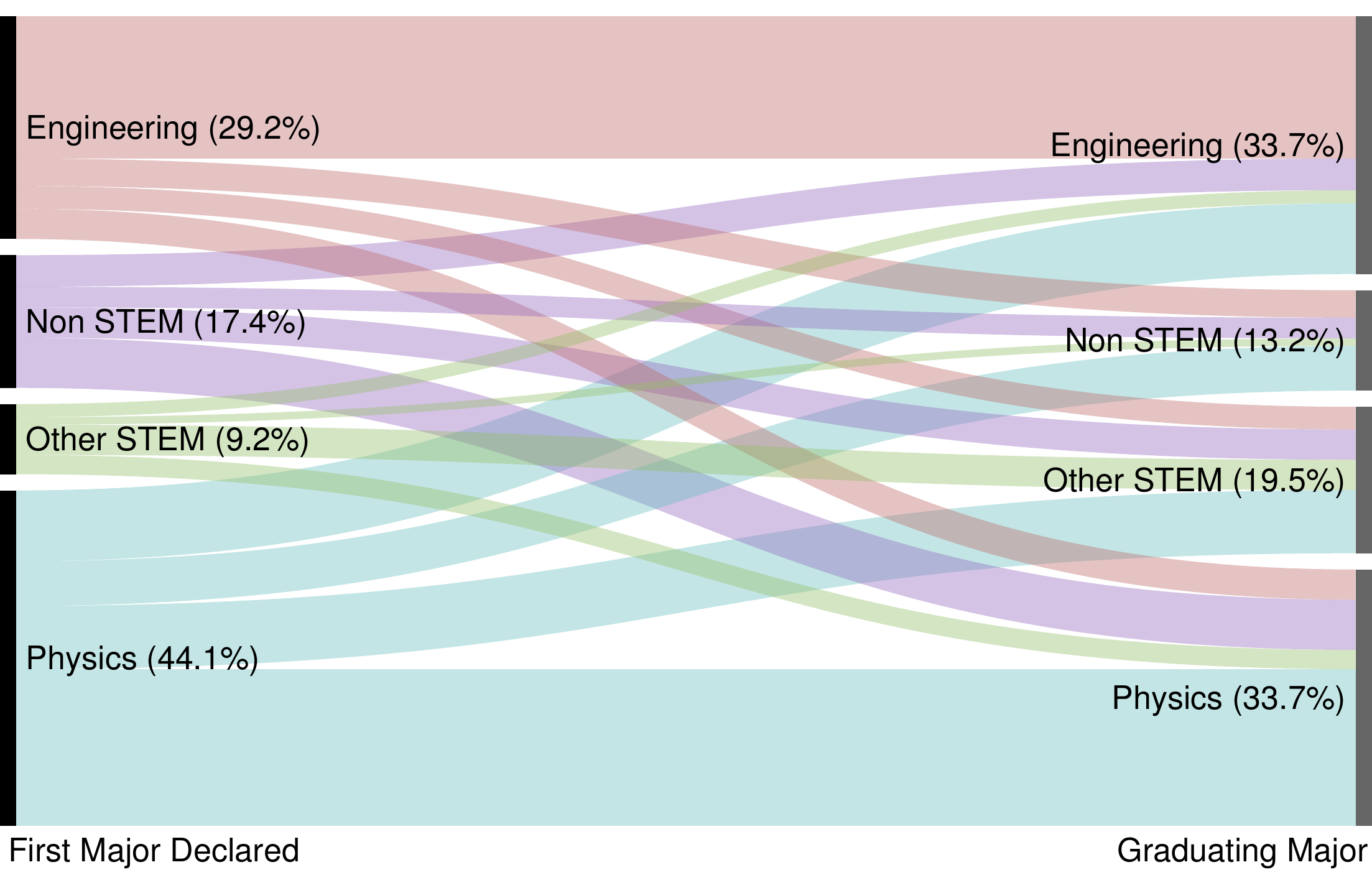}
\caption{[Color online] Approximately one third of students declaring a physics major go on to receive a degree in physics. Most students (87\%) who declare a physics major eventually receive a degree in STEM if they graduate. Groups on the left are the initial major declared by the student. Groups on the right are the graduating major the students receive a degree in.}\label{fig:alluvial}
\end{figure}

\begin{figure*}
\includegraphics[scale=0.35]{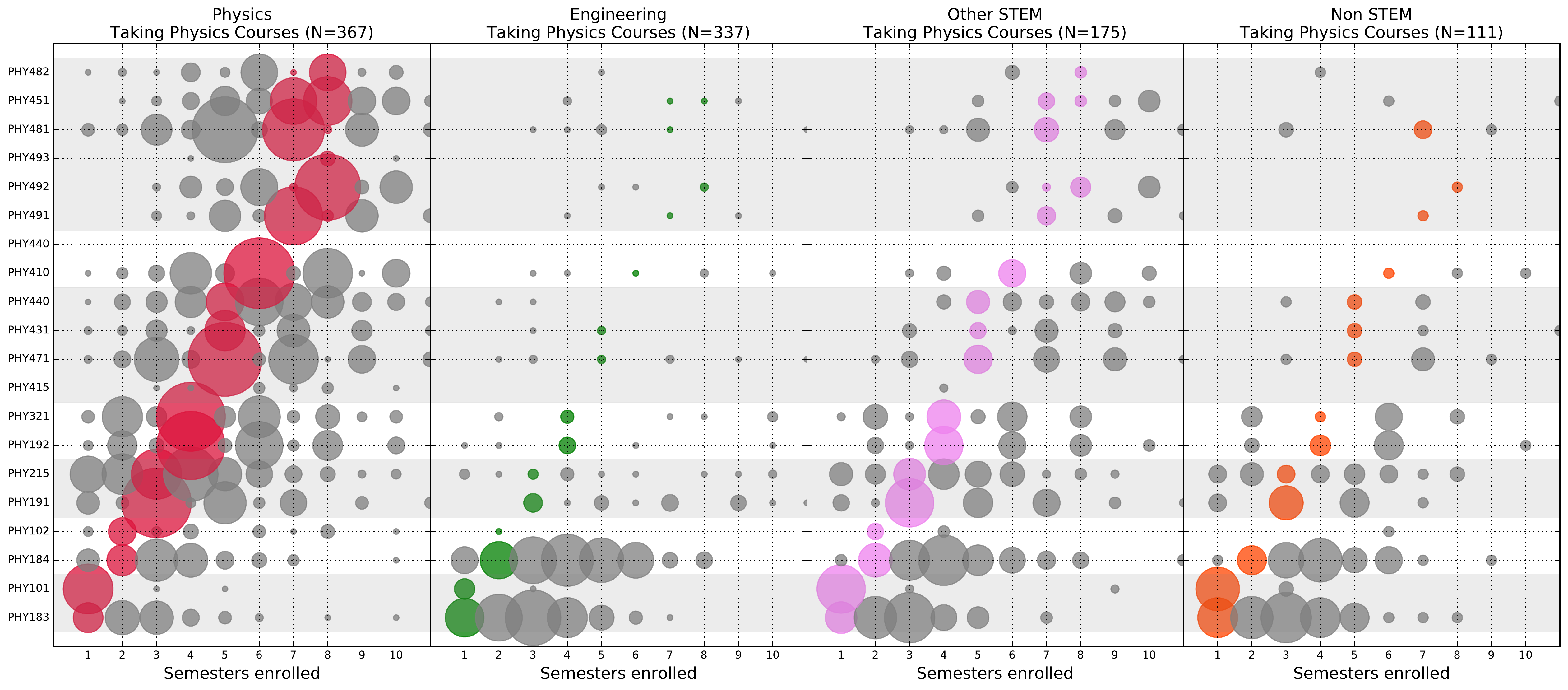}
\caption{[Color online] A time line of enrollment of students declaring a physics major separated by their eventual graduating degree. Non-physics graduates who have declared a physics major typically take physics after the recommended semester. Bubble size indicates the relative proportion of students taking the course in comparison to the entire group (i.e., physics, engineering, other STEM, non-STEM). Colored bubbles indicate courses taken during the recommended semester, gray bubbles indicate courses taken outside of the recommended semester, semester index is represented by the gray/white horizontal bars (first semester courses are at the bottom, senior level courses are at the top). Colors differentiate between exit degree obtained.}\label{fig:bubble}
\end{figure*}

We have begun to describe student pathways at two levels. One level looks at the starting major that students declare and the final major for which the student receives a degree. We visually represent the movement from start to finish using an alluvial diagram (FIG.~\ref{fig:alluvial}) \cite{alluvial}. This diagram helps visualize how student initial conditions affect graduation outcomes (e.g., what proportion of students graduate with their initially intended degree). A second level describes the order in which students complete each course required for the major. We represent this visually using a bubble diagram (FIG.~\ref{fig:bubble}). This level highlights the track that students take through the physics program and how completion of those courses relates to recommended, ``on-track'' schedule.

Approximately half (44.1\%) of students who declare a physics major at MSU do so when they first arrive at MSU. The remaining students switch into the physics major from a different degree programs or have not declared a major. Graduating students who declare a physics major are likely to remain in a STEM degree program even if they move away from physics (FIG.~\ref{fig:alluvial}). Approximately one-third (33.7\%) of students who attempt to get a degree in physics at MSU do so. An additional one-third (33.7\%) complete a degree in an engineering program.  The remaining students are likely to pursue other STEM offerings (e.g., chemistry or mathematics).

Students frequently complete physics courses outside of the recommended schedule by the physics department (FIG.~\ref{fig:bubble}). For example, students who declare a physics major but ultimately receive a different degree are most likely to take their introductory mechanics course (PHY 183) in their third semester and introductory electricity and magnetism course (PHY 184) in their fourth semester. Students who receive degrees in physics are more likely to take this introductory sequence prior to their third semester. Additionally, many students who eventually earn degrees in physics take senior-level E\&M (PHY 481) up to 1 year before the recommended schedule.  While we acknowledge there could be many reasons for students taking courses at different times, we are (currently) interested in finding useful representations that describe for a single institution what pathways students take through the major.


\section{Earned Grades differentiate physics graduates from others}\label{sec:grades}

In this initial study, we found that grades earned in math and physics courses differentiate students that eventually earn physics degrees from other graduates. Because course grades are not normalized measurements, we cannot compare raw grades between different courses, different course instructors, and different semesters. Thus, we have used the standard score or ``Z-score'' \cite{zscore} to normalize students' grades in a single course offering. 

\begin{equation}
    Z = \frac{x - \mu}{\sigma}
\end{equation}

The Z-score provides a measure of what fraction of a standard deviation ($\sigma$) that a particular value ($x$) deviates from the mean ($\mu$) of a distribution of scores. By using the Z-score, we do assume that students' scores within a given course offering are drawn from a normal distribution.

\begin{figure*}
\subfloat{\includegraphics[width=0.45\textwidth]{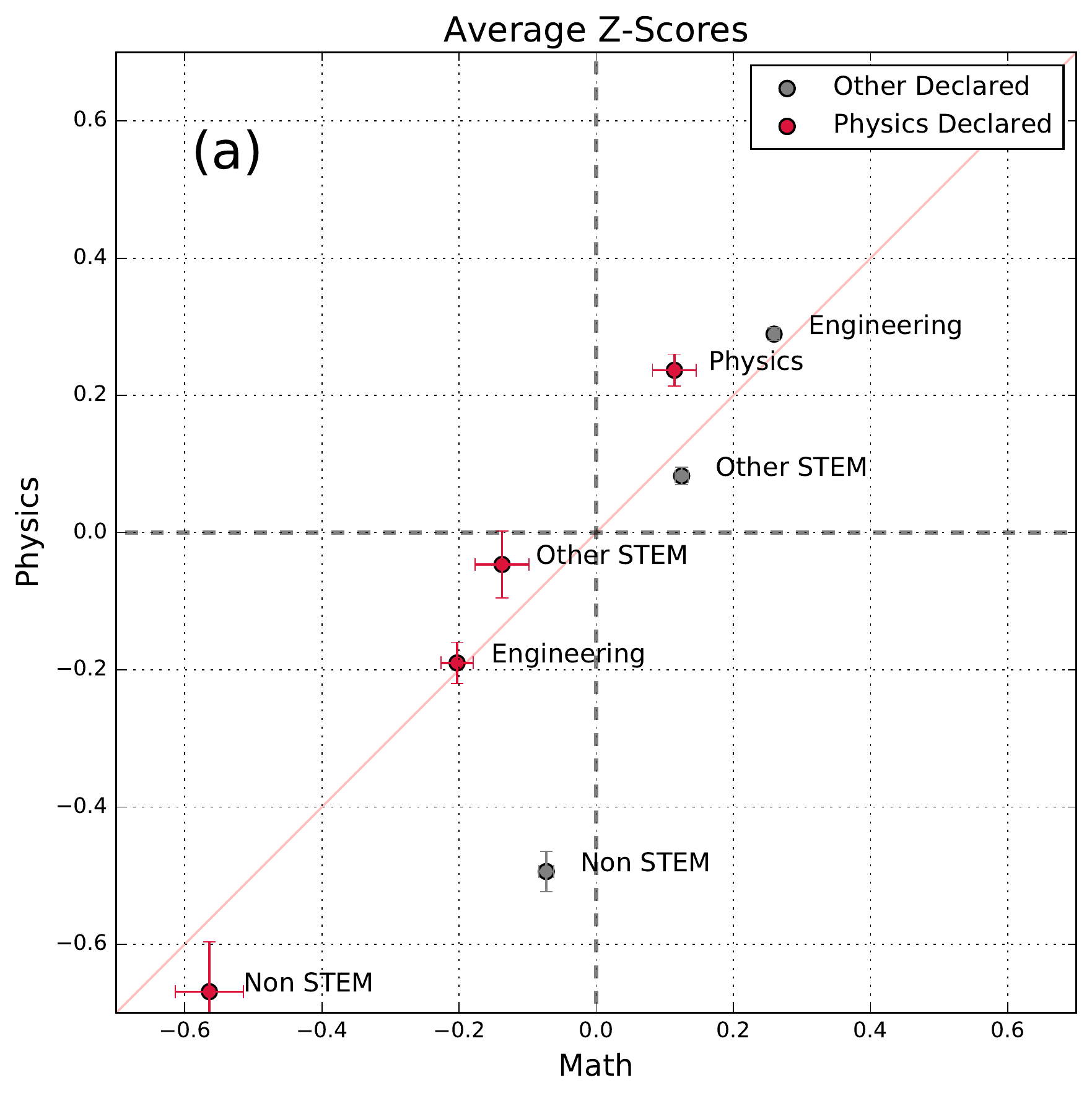}
}
\subfloat{\includegraphics[width=0.45\textwidth]{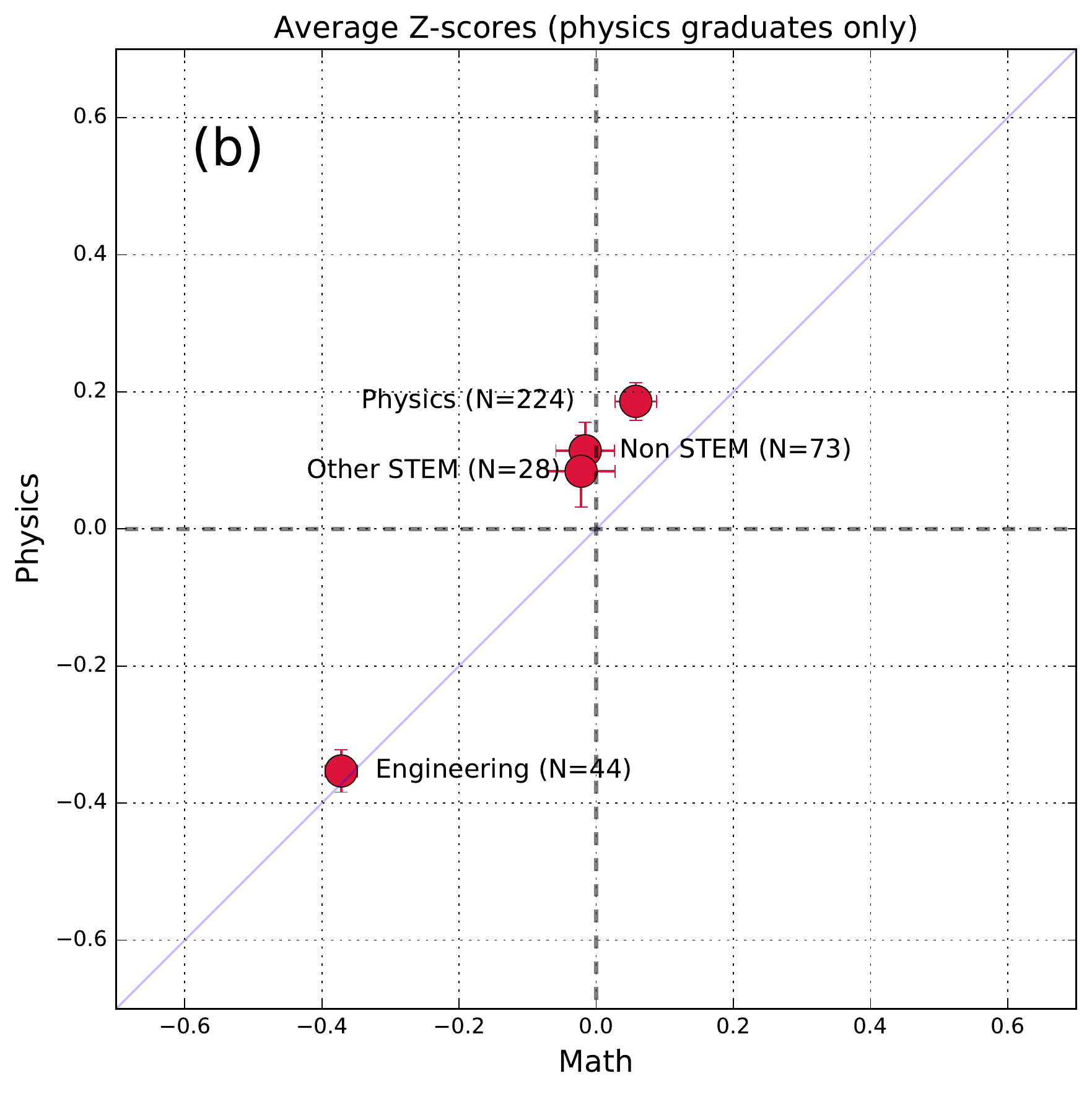}}
\caption{[Color online] Physics students receiving BS/BA degrees in physics/astronomy are above average in introductory course performance in comparison to students who move to different programs before graduating. (Fig \ref{fig:zscore}b) Students receiving physics degrees who initially declared engineering majors earn below average grades in introductory physics and math courses. The error bars represent the standard error of the mean for each axis. (Fig \ref{fig:zscore}a) data labels indicate degree category students received, (Fig \ref{fig:zscore}b) labels indicate initial major declared.
}\label{fig:zscore}
\end{figure*}

To compare groups of students, we first calculated the Z-score for each student in a particular course offering for every course from a restricted list of courses (described below). We grouped students by the degree they eventually earned (physics, engineering, STEM, and non-STEM) as well as by the condition of declaring physics as major any time in their academic career (yes, no) -- leading to 7 total groups. We then calculated the mean Z-score and standard error for physics courses and math courses separately for the population of students in each of the seven groups of students. These mean Z-scores along with their standard errors are shown in both in TAB.~\ref{tab:zscore} and FIG.~\ref{fig:zscore}.

We have restricted the courses from which we draw our data to only introductory courses (100 \& 200 level) required to earn a bachelor's degree in physics. Many different degree programs require these math and physics courses, thus they provide a large basis to compare students (TAB.~\ref{tab:zscore}). These courses include introductory mechanics (PHY 183), introductory electricity and magnetism (PHY 184), introductory lab courses (PHY 191, PHY 192), a third semester course covering thermodynamics and modern physics (PHY 215), and the calculus sequence from Calculus I to a first course in ordinary differential equations (MTH 132, MTH 133, MTH 234, MTH 235).

We find that students who receive a degree in physics perform above average in introductory math and physics (FIG.~\ref{fig:zscore}a; TAB.~\ref{tab:zscore}). We refer to these plots as ``normalized comparisons.'' Based on these normalized comparisons, students who declare a physics major but then move to other STEM programs/Engineering programs perform below average. Further, students who never declare a physics major and receive a degree in STEM/Engineering programs perform above average. We also find that students whose first declared major is an engineering program but ultimate degree is physics perform below average in physics and mathematics introductory courses.

\begin{table}[t]
\caption{Numbers of students and their corresponding normalized scores for the groups represented in FIG.~\ref{fig:zscore}a. Students are labeled by their graduating major and whether or not they ever declared a physics major.}
\label{tab:zscore}
\begin{tabular}{|l|l|l|l|}
\hline
Category (physics) & N & $Z_{math} \pm SE$    & $Z_{phys} \pm SE$\\ \hline \hline
Engineering (No)                            & 4047    & 0.26$\pm$0.01   & 0.29$\pm$0.01  \\ \hline
Non STEM (No)                               & 4913    & -0.07$\pm$0.01 & -0.49$\pm$0.03 \\ \hline
Other STEM (No)                             & 5833    & 0.13$\pm$0.01 & 0.08$\pm$0.01 \\ \hline
Engineering (Yes)                           & 374     & -0.20$\pm$0.02 & -0.19$\pm$0.03 \\ \hline
Non STEM (Yes)                              & 134     & -0.56$\pm$0.05 & -0.67$\pm$0.07 \\ \hline
Other STEM (Yes)                            & 202     & -0.14$\pm$0.04  & -0.05$\pm$0.05 \\ \hline
Physics (Yes)                               & 369     & 0.11$\pm$0.03 & 0.24$\pm$0.023 \\ \hline
\end{tabular}
\end{table}


\section{Discussion \& Conclusions}\label{sec:discussion}

In this paper, we have analyzed data collected by the registrar at MSU over the last 10+ years. This data can begin to provide information about the pathways that students take into and out of the physics major (for a given institutional context). In this methods paper, we have presented 3 representations (FIGS.~\ref{fig:alluvial}-\ref{fig:zscore}) that offer some shape to the story at MSU.

In particular, we have found that students earning physics degrees who have initially declared physics come from all areas of the university in roughly equal measure (FIG.~\ref{fig:alluvial}); contrary to departmental anecdotes. Students who leave the major also earn degrees in different areas in roughly equal measures, which is also is counter to the prevailing narrative in the department. Second, we find that students who earn physics degrees tend to follow the departmentally-recommended path up to the last year of their studies. We also find that students who eventually earn engineering degrees leave physics during or after the first E\&M course (PHY 184; Green circles in FIG.~\ref{fig:bubble}) while students who eventually earn other STEM degrees leave much later (Pink circles in FIG.~\ref{fig:bubble}). Finally, we find that students who are physics degree earners perform better in math and physics than students who declare physics and eventually earn some other degree, but perform not as well on course work than their engineering colleagues who never declared physics as a major (FIG.~\ref{fig:zscore}).

Through this work, we are not claiming that we have uncovered the full story from our current analysis or that all possible representations have been generated to explain our claims. Rather, we are suggesting that we have developed some methods and representations  (SEC.~\ref{sec:paths}) that provide some context for the paths that students take through the physics major at MSU as well as a possible mechanism for the observations of related to student attrition (SEC.~\ref{sec:grades}). While our results that show that students earning lower scores in math and physics courses are more likely to not earn degrees in physics (FIG.~\ref{fig:zscore}) are fairly obvious, we have also provided data that demonstrates that pathways of those degree earners are different from students earning degrees in other areas (FIG.~\ref{fig:bubble}).

The insight gained into the pathways that students take as gleaned from this data and our representations suggests there is a deeper and more interesting story that might exist in our data. For example, how do these pathways differ for different populations of students (e.g., based on incoming GPA, race, and ethnicity)? Furthermore, there are some analyses to be done that might provide additional context (e.g., how math and physics course enrollment and performance interact).

While our analyses and representations provide some context and detail about student pathways through the major, we recognize that by assuming a particular pathway for students to earn a degree (FIG.~\ref{fig:bubble}) that we are demphasizing alternative pathways and, likely, marginalizing non-traditional students. Moreover, that we assume a particular course trajectory for students to earn a physics degree might paint an unreasonably narrow picture of how students earn physics degrees. We are in the process of developing additional analyses that are not predicated on the student taking courses in a particular order. What we suspect is that a more comprehensive diagram that demonstrates the relationship between math and physics courses taken (i.e., in what order) will support our analysis and provide new and interesting information on student progression through the course work.

Finally, our present analysis neglects demographic information that might be important for understanding how different groups of students might be affected differently. As we construct new analyses and produce different representations of our data, we might find that asking similar questions of the data from this perspective will offer new insights into the pathways that women and under-represented students take through the major. Such an analysis is necessary if we are meant to foster and grow a diverse population of physics graduates.

\begin{center}
\textbf{ACKNOWLEDGMENTS}
\end{center}
This work was support by the College of Natural Sciences STEM Gateway Fellowship and Association of American Universities.












\end{document}